# Heuristic Classification of Physical Theories based on Quantum Correlations

**M. Ferrero**
Dpto. Física, Universidad de Oviedo. 33007 Oviedo, Spain.
maferrero@uniovi.es

**J.L. Sánchez-Gómez**
Dpto. Física Teórica, Universidad Autónoma de Madrid, 28049 Cantoblanco. Spain.
Jl.sanchezgomez@uam.es

**Keywords**: General Physics; Foundations of Quantum Mechanics; Locality; Contextuality; Classification of Physical Theories.

**PACS**: 01.30.Rr; 01.55.+b; 03.65.Ta; 01.70.+w;

**Abstract**

Taking quantum formalism as a point of reference and connection, we explore the various possibilities that arise in the construction of physical theories. Analyzing the distinct physical phenomena that each of them may describe, we introduce the different types of hidden variables theories that correspond to these physical phenomena. A hierarchical classification of the offered theories, based on the degree of correlation between dichotomic observables in bipartite systems, as quantified by a Bell type inequality, is finally proposed.

## 1.- Introduction.

In papers on foundations of quantum physics, and also in some on quantum information, it is frequent to read sentences which refer to the concepts of local hidden variables theories, contextual theories, non-signalling theories or non-local ones. However, the differences among these theories on the one hand, and their relationships with quantum mechanics on the other are not always clearly stated. Hence, many questions may arise. For instance: have local and non-contextual theories the same domain of application? Which one is able to elucidate more physical situations: a non-local (hidden) variables theory or a contextual one? Is quantum theory the most general type of theory? Is there a relationship between contextuality and non-signalling? These kinds of questions are usually overlooked and, consequently, they are not addressed in most of the relevant literature. In this paper we will try to address these issues qualitatively, filing things in folders so that younger generations of students, seeking a general understanding without excessive mathematical technicalities, need not to struggle again through all those notions. We will do it by sorting out the different types of theories introduced in the last 60 years according to their mutual relationship with the quantum mechanical formalism (QMF). The reason for this procedure is that, to compare physical theories, we need not only a common criterion, but also a basic mathematical framework. This will be the standard quantum theory, as described in the usual text books (for example, Cohen-Tannoudji et al, 1977, or Isham, 1995). In this essay, we will consider QMF as a physical theory empirically correct.

We use heuristic arguments to avoid the introduction of difficulties which would make the ideas obscure and hard to grasp by students, recent graduates in physics and philosophy coming to the field, philosophers of science and others scholars interested in the history and philosophy of quantum mechanics, for which the paper is mainly intended.

A certain quantity of what we say in the first part of the article might be known to researchers who are *especially* dedicated to the foundations of quantum physics. This is inevitable to put things in historical perspective, to introduce conceptual advances, and to get to the final classification. However, the substance of the essay is interwoven in the heuristic arguments that allow us to reach that classification, and not just on it.

A pedagogical survey like this always raises the question of whether or not it helps to illuminate the problems it raises. We will be satisfied if the reader understands the subtle differences between the theories; if we open the way to a more rigorous development of some of the ideas that are qualitatively introduced in this paper; if it helps to bring new light to old open problems; and if our preliminary results can afterwards be pushed forward to fill the gaps we left. That is, if its value is not alone in what we say, but also in what we may suggest.

The paper is organized with the following structure. First, and in order to have a reference frame, we remind the reader of some necessary basic quantum preliminaries. In part 3, the general idea of what a hidden variable theory is will be introduced. We will not give a historical account, since this would make the paper too long. The interested reader should consult in this respect different entries in the Stanford Encyclopaedia of Philosophy. An excellent book, but with the perspective of 40 years ago, is M. Jammer's book (Jammer, 1974). Instead, and by comparison with quantum formalism, we introduce Bell's formalism and go directly to the concept of *non-contextual* theories, the classical type of theories that have the most restricted range of application. This class will be presented then as a particular case of *local* theories. By contrast with the two previous cases, we bring in the *contextual* and nonlocal theories. Afterwards, we define the *non-signalling* theories and we discuss the non-local character of quantum formalism. Only at that moment we analyze some analogies and differences between the no-contextual, local, quantum, non-signalling, contextual and non-local theories. Part 4 contains a tentative classification of these theories based on the heuristic arguments of the previous discussion.

## 2. – Quantum Preliminaries.

In quantum mechanics, a physical magnitude $\mathcal{A}$ is represented by an observable $\hat{A}$. An observable is a hermitian operator that has a complete ensemble of eigenvectors $\{|v_\alpha\rangle\}$ on the Hilbert space $\mathcal{H}$. This complete ensemble can be discrete, continuous or both. In what follows, and to see better the connection with Bell formalism, we will considerer this basis as continuous in the real variable $\alpha$, which are the different values that the physical magnitude $\mathcal{A}$ can take. Complete means that: $\int d\alpha \, |v_\alpha\rangle\langle v_\alpha| = \hat{I}$, so that any quantum physical state $|\psi\rangle \in \mathcal{H}$ can be written as:

$$|\psi\rangle = \int d\alpha \, c(\alpha) \, |v_\alpha\rangle, \quad \text{where } c(\alpha) = \langle v_\alpha|\psi\rangle \quad (1).$$

To be an eigenvector $|v_\alpha\rangle$ of $\hat{A}$ means that it satisfies: $\hat{A}|v_\alpha\rangle = \alpha|v_\alpha\rangle$ (2).

Quantum theory *postulates* that in the *non-degenerate* (for simplicity) continuous case, if we measure the physical magnitude $\mathcal{A}$ when an individual system is prepared in the normalized state $|\psi\rangle$, the probability $d\mathcal{P}(\alpha)$ to obtain a result between $\alpha$ and $\alpha + d\alpha$ is given by:

$$d\mathcal{P}(\alpha) = |\langle v_\alpha|\psi\rangle|^2 \, d\alpha = \omega(\alpha) \, d\alpha, \quad \text{where } \omega(\alpha) = |\langle v_\alpha|\psi\rangle|^2 = |c(\alpha)|^2 \quad (3).$$

Suppose that we make N experiments with the system prepared always in the same normalized state $|\psi\rangle$, upon which we measure the physical magnitude $\mathcal{A}$. Let $d\mathcal{N}(\alpha)$ represent the number of cases in which we got a result between $\alpha$ and $\alpha + d\alpha$. When $N \to \infty$

$$\frac{d\mathcal{N}(\alpha)}{N} \to d\mathcal{P}(\alpha), \quad (4),$$

and, therefore, the average value of the magnitude $\mathcal{A}$ will be given by:

$$\frac{1}{N} \int \alpha \, d\mathcal{N}(\alpha), \quad (5)$$

which in the case in which $N \to \infty$ will tend to:

$$\left\langle \hat{A} \right\rangle_{|\psi\rangle} = \int \alpha \ d\mathcal{P}(\alpha) = \int \alpha \ \omega(\alpha) \ d\alpha \qquad (6),$$

where we have used the postulate (3).

To calculate the expectation value of any observable the usual rule is: $\left\langle \hat{A} \right\rangle_{|\psi\rangle} = \langle \psi | \hat{A} | \psi \rangle$. Using now that $\{|v_\alpha\rangle\}$ is a continuous basis on $\mathcal{H}$, and the relation (2), we can write:

$$\left\langle \hat{A} \right\rangle_{|\psi\rangle} = \langle \psi | \hat{A} | \psi \rangle = \int \alpha \ d\alpha \ \langle \psi | v_\alpha \rangle \langle v_\alpha | \psi \rangle = \int \alpha \ d\alpha \ |\langle \psi | v_\alpha \rangle|^2 = \int \alpha \ \omega(\alpha) \ d\alpha,$$

getting again, as it should be, expression (6).

Suppose now that $\mathcal{B}$ is another physical magnitude and that we measure it also N times and upon systems prepared in the same state $|\psi\rangle$. In the quantum formalism, $\mathcal{B}$ is represented by another hermitian operator $\hat{B}$ with a continuous basis on $\mathcal{H}, \{|u_\beta\rangle\}$, where again the $\beta$ are the different values that the observable $\hat{B}$ can take.

Let us imagine that we have a system prepared in some arbitrary state $|\psi\rangle$ and that we submit this system to three *successive measurements* of the physical magnitudes $\mathcal{A}$ and $\mathcal{B}$. In the first one we measure $\mathcal{A}$; in the second we measure $\mathcal{B}$, and in the third we measure $\mathcal{A}$ again. These measurements are performed in a quick succession in such a way that there is neither temporal evolution of the system between the first and the second measurement, nor between the second and the third.

*Definition*: Two physical magnitudes $\mathcal{A}$ and $\mathcal{B}$ are *compatible* if and only if the first and third outcomes coincides (Mandl, 1992. See also Peres, 1995). Therefore, compatibility means that the results of the measurements are independent of the order in which the measurements are performed (Jauch, 1968). For instance, in classical mechanics all pairs of observables are compatible.

*Theorem* (Gillespie, 1975): Given two physical magnitudes $\mathcal{A}$ and $\mathcal{B}$, represented by the observables $\hat{A}$ and $\hat{B}$, any of the three following conditions implies the other two:

. $\mathcal{A}$ and $\mathcal{B}$ are compatible.

. $\hat{A}$ and $\hat{B}$ have a common basis.

. $\hat{A}$ and $\hat{B}$ commute: $\left[ \hat{A}, \hat{B} \right] = 0$.

Suppose that $\mathcal{A}$ and $\mathcal{B}$ are compatible, and let $\{|\chi\rangle\}$ be their common basis. The average value of the product of the results obtained over many measurements of the commuting observables $\hat{A}$ and $\hat{B}$, can be calculated as follows:

$$\left\langle \hat{A} \hat{B} \right\rangle_{|\psi\rangle} = \langle \psi | \hat{A} \hat{B} | \psi \rangle =$$

$$\langle \psi | \left[ \int d\chi \ |\chi\rangle\langle\chi| \right] \hat{A} \left[ \int d\chi' \ |\chi'\rangle\langle\chi'| \right] \hat{B} \left[ \int |\chi''\rangle \langle\chi''| \ d\chi'' \right] | \psi \rangle = \int d\chi \ |\langle\psi|\chi\rangle|^2 \langle\chi|\hat{A}|\chi\rangle\langle\chi|\hat{B}|\chi\rangle \qquad (7),$$

where we have used the equality: $\langle\chi|\hat{A}|\chi'\rangle = \langle\chi|\hat{A}|\chi\rangle \ \delta(\chi-\chi')$ [similarly for $\hat{B}$] that follows from the fact that $\{|\chi\rangle\}$ are eigenvectors of both $\hat{A}$ and $\hat{B}$.

## 3.- Classes of H.V. Theories.

Historically, the motivation for introducing hidden variables theories was related to the lack of realism (in the restricted sense defined below) and determinism showed by the conventional interpretation of quantum mechanical formalism[1]. *Realism* in the broad sense means that there is a material reality, and in the restricted sense, that this material reality is composed by

---

[1] To be fear we should also add completeness and locality. See below.

*separable* objects that have dynamical intrinsic properties with *well-defined values* that *individualizes* them. A measurement of any property accomplished upon the quoted objects, correctly carried out with a suitably calibrated device, will reveal the pre-existing value of the intrinsic or underlying property

By *determinism* we understand the principle that declares that the effects are *uniquely and completely* determined by its causes (Ferrero et al., 2013). It is a well known fact that if one experiment is repeated with the system prepared always in the same initial state, different results can be obtained and, therefore, QMF does not satisfy the deterministic requirement.

One possible understanding of these difficulties with realism in the restricted sense and determinism was that, if for any reason beyond our actual capacities or experimental control in the *preparation*, the individual quantum systems making up the ensemble for the repetition, were *similarly* and not *identically* prepared, as intended, it would be possible to introduce additional parameters $\lambda$ able to explain the observed differences in the outcomes. This "completion" of the quantum formalism (Einstein et. al., 1935) would fully recover both realism and determinism, leaving open the possibility of a more detailed picture of the world that the one given by QMF. Those theories aimed to introduce these additional variables $\lambda$ not contemplated in the quantum formalism, are called, in general, *hidden variables theories* (HVT).

As a question of principle, we may think of two different classes of hidden variables theories. If the result of one observation of the physical magnitude $\mathcal{A}$, that we will denote by $a$ (a, $\lambda$), is always independent of any other measurement carried out over any other compatible physical magnitude of the system $\mathcal{B}$, $\mathcal{C}$, etc., the theory is called *non-contextual* (NC in what follows). Here the letter a is a parameter subject to experimental manipulation; for instance, in an EPRB experiment (Bohm, 1951), it specifies the spin component to be measured. These kinds of theories are fully realistic and deterministic in the sense defined above and, therefore, they clearly satisfy the motivation to introduce hidden variables.

On the other hand, if the result of measuring the magnitude $\mathcal{A}$ depends on which measurements of other compatible magnitudes $\mathcal{B}$, $\mathcal{C}$, $\mathcal{D}$, etc., are carried out, the theory is said to be *contextual* (C in what follows). We define the *context as the ensemble of compatible measurements that are realized*. This class of theories may seem now a bit strange from the point of view of restoring realism and determinism in physics. However, we will leave this possibility open.

Let us add, only for the sake of completeness, that a non-contextual hidden variable theory is *deterministic* if the result obtained in a measurement of a physical magnitude $\mathcal{A}$ depends only on the ontic variable $\lambda$ and on the adjustable parameter a of the apparatus. That is, if $a = a$ (a, $\lambda$) as already said.

A theory is *stochastic* if the result $a$ is also a function of another ensemble of parameters $\mu$ characterizing the environment in interaction with the measuring instrument: $a = a$ (a, $\lambda$, $\mu$). This case is similar to classical statistical mechanics, deterministic in principle, but impossible in practice due to the impossibility of controlling the variables $\mu$, which makes that the outcome can be only probabilistically predicted. It implies, thus, a certain degree of contextuality, because it would be possible to obtain different outcomes starting with the system (and the measuring device) in identical ontic states.

A theory is *essentially probabilistic* if the probability of obtaining the result $a$, conditioned to the system being in the state $\lambda$, the apparatus characterized by a, and the environment by $\mu$, is given by p = p (a, $\lambda$, $\mu$)[2].

It is not the purpose of this paper to discuss these last three distinctions in more detail. What is relevant is that starting from any of these three kinds of theories it would be possible to derive

---

[2] It is not very clear for us what this essential probability might mean in the context of HVT.

Bell inequalities (Bell, 87), so that nothing essential is lost by restricting oneself to the simplest case in which the underlying property determines the result (in the other two cases it is understood that the outcomes also depend on the intrinsic property, but the connection cannot be made explicit). Besides, the deterministic theories are also simpler from pedagogical point of view, so we will limit our arguments to them in which follows.

Let us suppose now that we prepare an ensemble of microscopic systems *identically* in a certain quantum state $|\psi\rangle$. By identically we meant that we reproduce in each round the same empirical procedures in order to guarantee that the same state is obtained. How this is achieved is a technical problem in which we cannot enter in this paper. We will assume that it can be assured that the state is the same.

Imagine that with half of these systems we perform an experiment, called experiment-1, in which we measure over each system one by one the physical magnitudes $\mathcal{A}$ and $\mathcal{B}$. The system may well be a bipartite system and, in that case, the experiment-1 could be imagined as one of the Aspect experiments (Aspect et al., 1982), but it may not need to be necessarily so. It does not matter now. If they are, we measure $\mathcal{A}$ on one side and $\mathcal{B}$ on the other. If not, we measure both magnitudes *locally* and upon the same system.

A (*deterministic*) *non-contextual hidden variables theory* (NC) is characterized as follows:

i) The state of an individual system is specified by the underlying property $\lambda$.

ii) The values taken by the properties $\lambda$ are distributed over an ensemble $\Lambda$ with a certain probability density $\rho(\lambda)$. This probability density satisfies:

$$0 \leq \rho(\lambda) \leq 1, \quad \text{normalized to} \quad \int_\Lambda \rho(\lambda)\,d\lambda = 1$$

iii) The values that a physical magnitude $\mathcal{A}$ can take are given by *a* fixed function $\mathcal{a}(a,\lambda)$, where $a$ is a parameter or a set of parameters controlled by the experimenter that characterizes the relevant degrees of freedom of the device measuring $\mathcal{A}$.

iv) The result of an individual observation of the physical magnitude $\mathcal{A}$ is independent of which other observations maybe made.

To reproduce the statistical predictions of quantum theory any NC theory needs to fulfil the following three conditions:

1st.- A one-to-one correspondence between the probability density $\rho(\lambda)$ and the states $|\psi\rangle$;

2nd.- A one-to-one correspondence between functions $\mathcal{a}(a,\lambda)$, $\mathcal{B}(b,\lambda)$, …etc, and quantum mechanical observables $\hat{A}$, $\hat{B}$,… etc. The range of $\mathcal{a}(a,\lambda)$ must be then the set of eigenvalues of $\hat{A}$,

3rd.- That the probability of getting a result $\alpha$ when measuring the physical magnitude $\mathcal{A}$, given by: $\int_{\Lambda(\alpha)} \rho(\lambda)\,d\lambda$, where $\Lambda(\alpha) = \{\lambda \mid \mathcal{a}(a,\lambda) = \alpha\}$, equals $|\langle v_\alpha|\psi\rangle|^2$.

With these definitions, the *hidden variable average* value for a physical magnitude $\mathcal{A}$ is given by:
$$\langle A \rangle_{nchv} = \int_\Lambda \mathcal{a}(a,\lambda)\rho(\lambda)\,d\lambda \qquad (8),$$

while the average value of the product of two observable magnitudes $\mathcal{A}$ and $\mathcal{B}$ is given by:
$$\langle AB \rangle_{nchv} = \int_\Lambda \mathcal{a}(a,\lambda)\,\mathcal{B}(b,\lambda)\,\rho(\lambda)\,d\lambda \qquad (9).$$

One of the necessary conditions for carrying out this program successfully is the existence of the one-to-one correspondence between operators and functions. The question to ask is thus: does a function $\mathcal{a}(a,\lambda)$ exist *for all the observables*?

The apparent trivial answer to this question is: yes, it does, for it you look at the two expressions (8) and (9) it seems trivial to see that they have the same form that the quantum mechanical expression (6) and (7) that we got previously, provided that we can make the following identifications: $a(a,\lambda) = \alpha$ ; $\rho(\lambda) = \omega(\alpha) = |\langle\psi|\chi\rangle|^2$ ; $\langle\chi|\hat{A}|\chi\rangle = a(a,\lambda)$ and $\langle\chi|\hat{B}|\chi\rangle = \mathcal{B}(b,\lambda)$. It seems intuitively obvious to think that this identification can always be done because $\alpha$ is a real parameter ($\hat{A}$ is hermitian), $\omega(\alpha)$ is a normalized probability density and the average values are real numbers.

However, the correct and non-trivial answer is: no. There is not a unique function $a(a,\lambda)$ for each observable. This non-trivial result was first envisaged by Gleason, in a corollary to his famous theorem (Gleason, 1957). Rediscovered by Specker (Specker, 1960) was finally re-rediscovered by John Bell in 1966 (Bell, 1987). Today it is usually known as the Bell-Kocher-Specker (BKS) theorem (Kochen-Specker, 1967). It is not the purpose of this essay to discuss these well-known theorems any further. We will only give an intuitive argument to show why a unique function $a(a,\lambda)$ cannot exist.

Imagine that over the other half left of the systems, identically prepared in the state $|\psi\rangle$, we perform another experiment, called experiment-2, in which we measure the physical magnitudes $\mathcal{A}$ and $\mathcal{C}$, and suppose also that the previous physical magnitude $\mathcal{B}$ is incompatible with $\mathcal{C}$. This implies that the hermitian operators that in the quantum formalism represents those magnitudes, the observables $\hat{B}$ and $\hat{C}$, do not commute and that $\{|\chi\rangle\}$ cannot longer be a common basis of the three: $\hat{A}$, $\hat{B}$ and $\hat{C}$. Let us then suppose that the new common basis for $\hat{A}$ and $\hat{C}$ is $\{|\eta\rangle\}$. Therefore $\{|\chi\rangle\} \neq \{|\eta\rangle\}$, necessarily. Using the quantum formalism, we will get a new expression for the average value of the observable $\hat{A}$ in this experiment-2:

$$\langle\hat{A}\rangle_{QM2} = \langle\psi|\hat{A}|\psi\rangle = \int d\eta \langle\psi|\eta\rangle \langle\eta|\hat{A}|\eta'\rangle \langle\eta'|\psi\rangle d\eta' = \int d\eta \ |\langle\psi|\eta\rangle|^2 \langle\eta|\hat{A}|\eta\rangle \qquad (10),$$

that has the structure of (8). Similarly, the average value of the product of the results obtained over many measurements of the compatible observables $\hat{A}$ and $\hat{C}$ would be given by:

$$\langle\hat{A}\hat{C}\rangle_{QM2} = \langle\psi|\hat{A}\hat{C}|\psi\rangle = \int d\eta \ |\langle\psi|\eta\rangle|^2 \langle\eta|\hat{A}|\eta\rangle\langle\eta|\hat{C}|\eta\rangle \qquad (11),$$

which also has the same form as (9). However, it should be already evident that in a non-contextual hidden variable theory, to get these same quantum results, we need to change the hidden parameters and therefore the functions $a$ from $a(a,\lambda)$ to $a(a,\zeta)$, let us say, in the two previous expressions. Consequently, if we wish to reproduce these quantum mechanical results, a *unique function* $a(a,\lambda)$ cannot exist. We have to change the function when we change the experiment perhaps because the function depends on the context, that is, on which other compatible observables are measured, (as happens with the QMF). The problem we have encountered may be understood in terms of the hidden variables program if we think that, in a localized *non-composite* microscopic system, the apparatus needed to measure $\mathcal{A}$ and $\mathcal{B}$ may be different from the apparatus needed to measure $\mathcal{A}$ and $\mathcal{C}$ (Bell, 1987). This sounds quite reasonable, although partially at variance with the motivation and the purpose of introducing hidden variables theories: if we need to change the hidden variable model when we pass from one experiment to the other, the measurement is not revealing anymore the *ontic properties* so cherished by the realist hidden variable program. Therefore, any hidden variables theory aiming at reproducing the statistical results of quantum mechanical formalism, must be a *contextual* one[3]. The BKS theorem disproves the existence of no-contextual hidden variables theories. Furthermore, these theoretical advances have been empirically confirmed very recently. It has been shown that any non-contextual hidden variable model is *experimentally* excluded (Lapkiewick et al., 2011).

---

[3] Contrary to other authors (Svozil, 2013) we consider that contextuality in quantum mechanics has its origins in the projection postulate and is already present in the case of a three level system in which two commuting observables are measured.

John Bell went much further, reaching another very important non-trivial result. He considered that if the initial identically prepared systems were bipartite systems, that is, *composed* by two parts 1 and 2 that we suppose now very far away, any *spatially separated* measure carried out upon the subsystem 2, cannot have any effects upon the subsystem 1 (*locality principle*). In particular, imagine that in one experiment we measure the dichotomic[4] (values +1 and -1) physical magnitudes, $\mathcal{A}$ over the subsystems 1 and $\mathcal{B}$ upon 2, and that in another experiment we measure $\mathcal{A}$ upon 1 (without introducing any change whatsoever with respect to the previous case) and $\mathcal{C}$ upon 2. Besides, let us suppose that the dichotomic observables $\hat{A}$ and $\hat{C}$ are not compatible. In these circumstances, the locality condition, Einstein's great worry, states that there are not *a priori* physical reasons to change the function $\mathcal{A}$. The non-trivial result obtained by Bell was that the average values[5] of the product of the outcome of one observation on subsystem 1 with the outcome of an observation on subsystem 2 satisfies an inequality, called the Bell inequality, and that there are states for which this inequality is violated by the predictions of the quantum mechanical formalism (Bell theorem). Therefore, he proved that, even if the subsystems are spatially separated, *we still need to change the function $\mathcal{A}$* and that there exist quantum states that do not satisfy the locality condition. Hence, any hidden variable theory aiming at reproducing all the statistical results of quantum theory must be *non-local*. Note that in the previous discussion non-locality may be seen as a particular case of contextuality.

This non-locality of HVT may be also explained in a way similar to which we did above to introduce the BKS theorem. In the standard quantum mechanical formalism, there are states for which the average is:

$$\left\langle \hat{A}\,\hat{B} \right\rangle_{QM} = \left\langle \psi \middle| \hat{A}\hat{B} \middle| \psi \right\rangle = -\vec{a}\cdot\vec{b} \qquad (12),$$

Bell's question was: Can we find functions $\mathcal{A}$ and $\mathcal{B}$ and a probability distribution $\rho(\lambda)$ to reproduce this correlation (12)? The answer he found was yes. There are many of such functions with the form: $\mathcal{A}(a,b,\lambda)$, $\mathcal{B}(a,b,\lambda)$ and $\rho = \rho(a,b,\lambda)$. However, if we add now the *locality condition*, that amounts to say that on the one hand the setting b (or a) of a particular instrument has no effect on what happens in a spatially separated region, requisite called parameter independence (PI), plus, on the other hand, that all the correlation between the results have to come from the common past[6], requisite called outcome independence (OI), that is if we want to find functions such that: $\mathcal{A} = \mathcal{A}(a,\lambda)$, $\mathcal{B} = \mathcal{B}(b,\lambda)$ and $\rho = \rho(\lambda)$, Bell found that in general such functions do not exist[7].

From the point of view of the realist philosophy behind the hidden variables program, the two previous conclusions are very inconvenient, to say the least. The world in which we live seems to be a classical world, where intrinsic properties do exist, separate subsystems behave individually, and non-local actions do not exist. At this level and in all theories of science, modulo QM, local realism is an excellent assumption. However, these two non-trivial results show that if one manages to construct a "classical type" of theory able to reproduce all quantum

---

[4] Observe that any physical magnitude may be made dichotomic.
[5] These average values measure *the correlation* between the corresponding physical magnitudes. In effect, the correlation coefficient may be defined by:

$r(\mathcal{A},\mathcal{B}) = \dfrac{\langle AB \rangle - \langle A \rangle \langle B \rangle}{\sigma_A \cdot \sigma_B}$, were $\sigma_A = \left( \langle A^2 \rangle - \langle A \rangle^2 \right)^{1/2}$ and the same for B. In the case of dichotomic physical magnitudes, obtained with equal probability, $\langle A \rangle = 0 = \langle B \rangle$ and $\sigma_A = 1 = \sigma_B$, so now $r(\mathcal{A},\mathcal{B}) = \langle AB \rangle$.

[6] Note that if they are spatially separated, hidden communication is excluded.
[7] QMF satisfies parameter independence, but non-separable quantum states may violate outcome independence. See Jarret, 1984 and Shimony, 1993. More recent discussions of this topic can be also seen in Maudlin, 2011.

mechanical predictions, that hypothetical theory must be contextual and non-local in the sense stated above. Joined together they demolish, in one way or another, the classical picture of reality so close to the hidden variables program.

To clarify any misunderstanding about what we have said and for pedagogical reasons, a relevant remark is at order. It is frequently stated that Bell's theorem shows that quantum theory is non-local. In our opinion, this is still a controversial issue. To be sure, the only conclusion so far reached in this respect is that if we want to build up a hidden variable theory able to reproduce the statistical predictions of quantum theory, *this* HVT must be non-local. Following Bohr, d'Espagnat and other authors, this is the sense in which we will always use the term non-locality. To put it another way, in the context of the QMF it seems to us more appropriate to speak about entangled states and non-separability. Note that the two concepts are no equivalent. In QM there are non-separable states, for instance, Werner states (Werner, 1989) that do not violate Bell inequality. They do not violate either PI, or OI.

Let us go a step further by considering in what follows bipartite systems and observables with two possible results, like in Aspect´s experiments. As in the previous experiments, on one side Alice measures by order the magnitude $\mathcal{A}$ (represented in QMF by the observable $\hat{A}$) upon half of the subsystems of the ensemble at her disposal, and on the other side, very far away, Bob measures upon the corresponding pair the magnitude $\mathcal{B}$. In another experiment Alice measures, by order and over her halt left, $\mathcal{A}$ and Bob $\mathcal{C}$ ($\mathcal{B}$ and $\mathcal{C}$ are represented by the non-commuting observables $\hat{B}$ and $\hat{C}$) also by order and upon the corresponding pair. After finishing these two experiments, Alice calculates the averages $\langle A \rangle_1$ and $\langle A \rangle_2$. $\langle A \rangle_1$ is the average value for the magnitude $\mathcal{A}$ when the dichotomous magnitudes $\mathcal{A}$ and $\mathcal{B}$ are measured, and $\langle A \rangle_2$ is the average value when the magnitudes $\mathcal{A}$ and $\mathcal{C}$ are measured.

**Definition:** A theory is a *non-signalling theory* if and only if in the stated circumstances $\langle A \rangle_1 = \langle A \rangle_2$, that is, if the average value obtained by Alice is independent of the operations performed and the outcomes obtained by Bob. It is trivial to see, by a simple inspection of the expressions (6) or (10), that quantum formalism satisfies this condition: $\langle \hat{A} \rangle_{QM} = \langle \psi | \hat{A} | \psi \rangle$ and, therefore, it is a non-signalling theory. Had it been otherwise, that is, that $\langle A \rangle_1 \neq \langle A \rangle_2$, they could use this difference to send instantaneous messages (*signalling*)[8]. So far, it has been impossible to contrive an experiment to do that, and we have good reasons, namely relativity theory, to think that it cannot be done. Theories that do not allow sending information at superluminal velocity are called non-signalling (NS) theories. Therefore, in these classes of theories, $\langle A \rangle_1 = \langle A \rangle_2$, independently of which other operations are done by Bob on the other side.

Observe that if Bob tells Alice, one by one, the results that he is getting, this classical information would allow her to choose a sub-ensemble of systems on which $\langle A \rangle_1 \neq \langle A \rangle_2$. In these circumstances, she will get the false impression that Bob is steering the states of the subsystems at her disposal. He is not. Alice own intervention is essential (through the mentioned post-selection process) to allow this non-local presentation of QMF. However, it is clear that, in these circumstances the interchange of information is not instantaneous.

**4. - A tentative classification of physical theories.**

In the previous sections, we have introduced the following different classes of theories: quantum formalism; (deterministic) non-contextual and contextual hidden variables theories; local and non-local theories; and non-signalling theories. It seems natural to ask now if there are hierarchical relationships between them.

---

[8] Any theory that violates parameter independence would be a signalling theory (Ferrero, et al., 1997).

In what follows, we will develop qualitatively these possible relationships at the appropriate level at which this article is intended. To describe with mathematical rigor the relations among these classes of theories is really an open hard question that falls well beyond the scope of a pedagogical paper. The interested reader may consult technical discussions in Brandenburger et. al., 2008, (which are focussed in the mathematical properties of the HV theories). Here we will simplify considerably the problem by considering the ensemble of the distinct physical situations that each class of theories may elucidate and with reference to the simplest experimental case mentioned above, namely that in which in each run two observables with two possible outcomes are measured in bipartite systems. Nothing is lost if we think that one of the parties is in Alice's side, and she measures two dichotomic observables, and the other party is in Bob's side, and he also measures two dichotomic observables.

As we have explained, local and non-contextual theories do not coincide and they maintain an inclusion relationship that we will discuss now. A non-contextual theory is always, by definition (see part 3 above) also local, although the converse is not fulfilled. A local theory may be contrived to describe physical situations that are contextual. It would be enough that the results obtained on one side being independent of any operation made on the other side, but they still might depend on which other compatible observables are measured on its side, that is, locally. On the other hand, situations in which a local theory may become a non-contextual theory may also be contrived. This transformation would be intuitively easy to understand. Suppose that we have a system composed of two subsystems. If the theory is local we have to get the same results when the subsystems are spatially separated and if we gradually bring them closer until they are together. The situation finally reached would correspond to a local and non-contextual one (provided that the outcomes are independent).

We conclude thus that local theories are the more general of the two and that they include properly the non-contextual ones. In other words, local theories explain a bigger ensemble of physical phenomena than the non-contextual ones, as represented in figure 1.

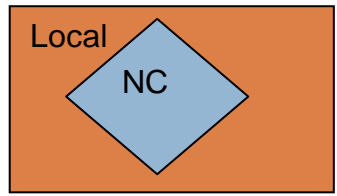

Figure 1. Ensemble of physical phenomena describable by non-contextual and local theories respectively.

The regions in the figure correspond to physical phenomena that may be describable with a non-contextual model or with a local one. A point in the NC region may be related to a set of equivalent experimental procedures. The inclusion is proper, so any conceivable test explainable by a NC model will be also explainable by a local one, but, as we have discussed above, the converse is not true. The degree of generality increases, thereby, with the surface of the region in the figure. The region outside NC is the complementary of it, that is, it represents the set of phenomena describable by contextual theories. For instance, all physical phenomena describable by classical physics lie in the region L.

Let us now see what happens with quantum theory. Are all experimental test described by QMF describable by a local HVT? As previously discussed in this article, the answer to this question is a clear no. Quantum correlations are not explainable using local models. However, any classical physical situation may be understood using quantum formalism. Separable quantum states can be used to explain experimental test lying in the region L. Consequently, QMF is able to describe a bigger set of physical phenomena than any local HVT, the inclusion being also proper.

A first attempt to classify these three classes of theories taking as criterion the set of physical phenomena that they are able to describe would be:

$$NC \subseteq Local \subseteq QM \qquad (13).$$

In the previous pages we have also spoken about another three types of theories, which we will intend now to include in our tentative classification: NS, Contextual and NL. Let us say from the outset that the relationships between these theories and the previous ones introduced are intuitively not very clear. To be able to advance without introducing too many difficulties, we will simplify the problem by considering the degree of correlation between dichotomic observables in bipartite systems, *as* quantified by the Clauser-Horne-Shimony and Holt (CHSH) inequality (Clauser et al., 1969). This will allow us to propose a new relationship that will be useful to establish the classification we are looking for. The CHSH is a Bell type inequality satisfied by any local theory that can be written as: $S \leq 2$, where S is a certain observable. Quantum mechanics, on the other hand, is a no-signalling theory that violates Bell inequality until $S \leq 2\sqrt{2}$. That is, it may reach a higher degree of correlation than the local ones. The reason why *properties that do not existed* before the measurement, as the conventional interpretation of QMF maintains, reach a degree of correlation higher than the actual ontic properties of LHV is a surprising result, especially from the classical realist point of view. It is hard to imagine the physical reason behind the fact that "the creation of properties" may lead to stronger correlations than "the revelation of properties", a very counterintuitive insight.

Although somewhat artificial, there are non-local models, called PR boxes or non-local machines by Popescu and Rohrlich, (Popescu and Rohrlich,1994) -previously noted by Tsirelson, (Tsirelson,1980)-, that violate the CHSH inequality until S = 4 *without violating the no-signalling constrain*. This clearly shows that using non-signalling models it is possible to reach a higher degree of correlation than in quantum mechanics and, therefore, that they are able to describe physical phenomena that are not possible to describe with QMF. The reason why these stronger correlations cannot be attained by quantum formalism is in our opinion still an open problem not well understood. To go beyond this limit is not easy. For example, is there a connection between contextuality and no-signalling? We do not have an answer to this question. However, QMF has these two characteristics, so there might be formal relationships between them that it should be interesting to investigate.

Although brief, our heuristic discussion, in which the degree of correlation attainable has played the key role, would allow us to extend (13) until the following relation:

$$NC \subseteq Local \subseteq QM \subseteq NS \subseteq Signalling (NL/C).$$

Even though we have been able to include one region inside other with a greater degree of generality, many aspects still remain obscure, in particular those related to the outermost region, where NL, contextuality and signalling coexist. It is difficult to advance here, even with qualitative arguments, because we do not have criteria to sort out these three types of "theories". All we may have are questions. Consequently, we will limit our analysis to the relationship between contextuality and non-locality, leaving signalling for our final comment.

As we have already discussed, all non-local theories are, by definition, contextual, but the reciprocal is incorrect: not all contextual theories are non-local. As we have say above, QMF may be presented as a contextual and local theory: if a hidden variable theory wants to reproduce quantum predictions obtained using some non-separable states, this HVT must be non-local. In any case, the so-called quantum non-locality restrings to non-separable states, whereas contextuality can be experimentally observed for any state of any non-composite system with at least two compatible observables (Lapkiewicz et al., 2011). Quantum correlations do not need parts, and therefore contextuality appears as being able to describe more general phenomena than non-locality (Cabello 2011). Besides, it seems intuitively easy to conceive, at least hypothetically, cases in which an experimental situation explainable by a local and contextual model could be transformed into a non-local one and reciprocally. Suppose that we have a contextual model built up to explain locally certain experiment. Now, if these two subsystems are in turn composed, we could transform this local contextuality type of situation into a non-local one by physically separating its parts (Cabello, 2010). To go in the opposite direction is perfectly conceivable: we can transform a non-local type of situation in a local but contextual one by joining its parts. Therefore, our guess is that the contextual theories are the more general of the two.

The above heuristic considerations suggest a hierarchical classification of the different theories introduced for the considered case as represented in the figure 2 bellow.

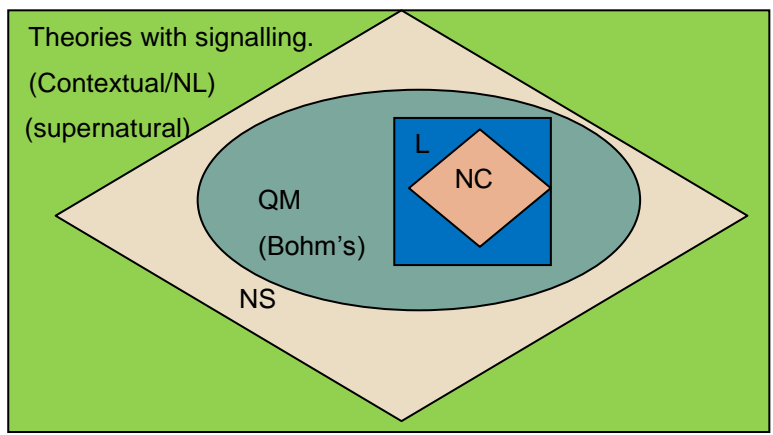

Figure 2. Regions representing physical phenomena describable by the distinct types of theories considered in the text.

This figure is a natural continuation of figure 1. Here NL is in fact the complementary of L, Contextual the complementary of NC, and although the figure does not show it up, they cover the entire area outside NC and L respectively. This implies that any point outside NC corresponds to physical phenomena whose explanation requires a contextual theory, and the points outside L corresponds to physical phenomena whose explanation requires a NLHV theory. For instance, Relativity theory is a NS theory describing phenomena which lie inside L, and the Aspect correlation experiments lie inside the QM part of the diagram: their results cannot be explained neither by a local model, nor by a non-contextual one. Outside NS any possible theory would require signalling.

If, as we have said, the degree of generality increases with the surface of the depicted region, our figure indicates that contextuality is more general than non-locality, as previously conjectured.

Notice here that the figure also suggests the standard path to unify QM and relativity theory. If the latter theory is placed inside the L region, as any classical (and local) theory should, it would indicate that QM is a more general type of theory and therefore, that the way forward to unify them would be to quantize gravity.

A brief philosophical final comment seems to be now necessary. As we have said, the region NC fully coincides with common sense and also with classical realism. Therefore, going from this region to the external part of the figure, the degree of classical realism gradually decreases and the difficulties with our intuitive understanding increases. In the following region L, there are physical situations locally contextual. Although still inside the classical realist program, not all the properties can be now understood merely as intrinsic or ontic properties describing an underlying reality.

Next region corresponds to QM. Here we have non-separable states that may describe situations that admit stronger correlations than the ones allowed by any local model. Intrinsic properties do not exist in general and the outcomes are not deterministically predictable. Realism in the restricted sense is invalidated. However, in the broad sense it can be maintained, although not as something that exists independently of what is experimentally established[9].

The NS region of the figure remains practically unexplored at the theoretical level, and there is little more to add to what we have already said. And with regard to the outermost one, signalling appears to be devoid of scientific interest. It seems evident that if we allow the outcomes of the experiments to depend instantaneously on all the settings, everything could be explained. It does not exclude any physical situation and, therefore, it says nothing about how material reality can be. In this region would fall all those theories that are more close to the supernatural explanations than to science.

**Acknowledgements.**
We thank A. Cabello and A. Cassinello for fruitful discussions. We would also like to thank E. Santos, R. Lapiedra and V. Gómez-Pin for reading the manuscript and for helpful suggestions.

---

[9] We should review now that there is at least one HVT that is deterministic, contextual and non-local without allowing signalling (Bohm's theory). But this fact does not alter in any way the established relation of inclusion.